\documentclass[a4paper,10pt]{revtex4}
\usepackage[utf8]{inputenc}
\usepackage{amsmath}

\begin{document}
\author{Yamen Hamdouni}
\email{hamdouniyamen@gmail.com}
\affiliation{Department of physics, Faculty of Exact Sciences, Mentouri University, Constantine, Algeria}
\title{Exact lower bound of the uncertainty principle product for the harmonic oscillator with position-momentum coupling }
\begin{abstract}
We show that the uncertainty principle product for the position and momentum operators for a system described by the Hamiltonian $ \hat H= \frac{\hat{p}^2}{2m} +\frac{1}{2} m \omega^2 \hat{x}^2+\frac{\mu}{2}(\hat x \hat p+ \hat p \hat x)$ where $\mu<\omega$ reads  $\Delta x \Delta p\ge\frac{\hbar \omega}{2\sqrt{\omega^2-\mu^2}}$. All the values bellow this lower bound are thus quantum-mechanically forbidden. We construct the annihilation and creation operators for this system and we calculate the expectation values of the operators $\hat p$ and $\hat x$ with respect to the corresponding coherent states.
\end{abstract}
\maketitle

\section{Hamiltonian and wave functions}
The Heisenberg uncertainty principle implies that the uncertainties on the canonically conjugate operators $\hat x$ and $\hat p$ satisfy the general inequality~\cite{heis}
\begin{equation}
 \Delta x \Delta p\ge\frac{\hbar}{2}.
\end{equation}
It is natural to question the possibility of the determination of the exact lower bound of the uncertainty product for particular physical systems that are of fundamental interst. In what follows we shall be mainly interested in a harmonic oscillator whose Hamiltonian is of the form
\begin{equation}
\hat H= \frac{\hat{p}^2}{2m} +\frac{1}{2} m \omega^2 \hat{x}^2+\frac{\mu}{2}(\hat x \hat p+ \hat p \hat x).\label{ham}
\end{equation}
The time-independent Schroedinger equation associated with Hamiltonian (\ref{ham}) reads:
\begin{equation}
 -\frac{\hbar^2}{2m} \frac{d^2 \psi(x)}{dx^2}+\frac{1}{2}m\omega^2 x^2 \psi(x)-\frac{i\hbar \mu}{2}\Bigl(2 x \frac{d\psi(x)}{dx}+ \psi(x)\Bigr)=E\psi(x)\label{shro1}.
 \end{equation}
By inspecting the behavior of the above equation when $x\to\pm\infty$ where the wave function should vanish (finite motion), it turns out that its solution may be written as
\begin{equation}
 \psi(x)=e^{- \rho x^2} \xi(x)\label{tran}
\end{equation}
where $\rho$ is a complex number whose real part is positive  ($\rm Re \rho >0$) and $\xi(x)$ is a well-behaved function.

Inserting the solution (\ref{tran}) into the Schroedinger equation (\ref{shro1}), we obtain:
\begin{eqnarray}
 && -\hbar^2 \frac{d^2 \xi(x)}{dx^2}+(2\hbar^2\rho-i\hbar m\mu) x\frac{d\xi(x)}{dx}\nonumber \\
  &+&\Bigl(2 \hbar^2 \rho(1-2\rho x^2) + i\hbar m(4\rho x^2-1)+m^2\omega^2 x^2\Bigr)\xi(x)=2mE\xi(x)\label{shro2}.
\end{eqnarray}
Let us choose the parameter $\rho$ such that the coefficient of $x^2$ in equation (\ref{shro2}) vanishes, namely:
\begin{equation}
 4\hbar^2\rho^2-4i\hbar\mu m\rho-m^2 \omega^2=0.
\end{equation}
This gives:
\begin{equation}
 \rho=\frac{m}{2\hbar}(i\mu\pm\sqrt{\omega^2-\mu^2}).
\end{equation}
In order to satisfy the requirements cited earlier regarding the behavior of the wave functions at $\pm\infty$, the only acceptable choice for $\rho$ is
\begin{equation}
 \rho=\frac{m}{2\hbar}(i\mu+\sqrt{\omega^2-\mu^2}), \qquad \mu<\omega.
 \end{equation}
 The condition $\mu<\omega$ is thus necessary for the motion to be finite. With this choice, the differential equation for the function $\xi(x)$ simplifies further to become
 \begin{equation}
  \xi''(x)+\frac{ m \sqrt{\omega^2-\mu^2}}{\hbar} \Biggl[\biggl(\frac{E}{\hbar \sqrt{\omega^2-\mu^2}}-\frac{1}{2}\biggr)\xi(x)-x \xi'(x)\Biggr]=0.
 \end{equation}
Setting 
\begin{equation}
 n=\frac{E}{\hbar \sqrt{\omega^2-\mu^2}}-\frac{1}{2}\label{ener}
\end{equation}
 leads to
 \begin{equation}
  \xi''(x)+\frac{ m \sqrt{\omega^2-\mu^2}}{\hbar} (n\xi(x)-x \xi'(x))=0,
 \end{equation}
 which is the differential equation of a Hermite's function, and thus its solution is
 \begin{equation}
  \xi_n(x)=C_n H_n\bigl(\sqrt{\tfrac{\hbar}{m}}(\omega^2-\mu^2)^{\frac{1}{4}} x\bigr),
 \end{equation}
where $n$ is a positive integer. Thus the full solution of the Schroedinger equation reads:
\begin{equation}
 \psi_n(x)=\frac{1}{\sqrt{2^n n!}} \Biggl(\frac{m \sqrt{\omega^2-\mu^2}}{\pi \hbar}\Biggr)^{\frac{1}{4}} e^{-\frac{m}{2\hbar}(i\mu+\sqrt{\omega^2-\mu^2})x^2} H_n\bigl(\sqrt{\tfrac{\hbar}{m}}(\omega^2-\mu^2)^{\frac{1}{4}} x\bigr),
\end{equation}
where we have used the orthogonality relation of Hermite's functions to determine the constant $C_n$. From equation (\ref{ener}), it follows that the energy  of the system is quantize in the sens that:
\begin{equation}
 E_n=\hbar \sqrt{\omega^2-\mu^2} (n+\frac{1}{2}), \qquad n=0,1,2 \cdots.
\end{equation}

\section{Constructing the annihilation and creation operators}
We see that  the wave function $\psi_n(x)$ is equivalent to that of a one dimensional harmonic oscillator with frequency $\sqrt{\omega^2-\mu^2} $.  As we shall demonstrate bellow,  eventhough   the phase factor $ e^{-\frac{m}{2\hbar} i\mu x^2}$ does not affect  the modulus of the wave function, it plays, however, a crucial role as far as quantum fluctuations and uncertainties in position and momentum are concerned. 

For the time being, we focus on deriving the creation and annihilation operators for the aforementioned oscillator. To this end, we remark that the ground state wave function reads:
\begin{equation}
 \psi_0(x)= \Biggl(\frac{m \sqrt{\omega^2-\mu^2}}{\pi \hbar}\Biggr)^{\frac{1}{4}} e^{-\frac{m}{2\hbar}(i\mu+\sqrt{\omega^2-\mu^2})x^2}.
\end{equation}
We write the annihilation operator as a linear combination of the position and momentum operators:
\begin{equation}
 \hat a=i(\gamma \hat x+\theta \hat p),
\end{equation}
where $\gamma$ and $\theta$ are complex numbers satisfying the condition:

\begin{equation}
 2\hbar \rm Im (\theta \gamma^*)=1. \label{norm}
\end{equation}
 Since the operator $\hat a$ annihilates the ground state, then:
\begin{equation}
 \Bigl(\gamma  x-i\hbar \theta \frac{d}{dx}\Bigr)  e^{-\frac{m}{2\hbar}(i\mu+\sqrt{\omega^2-\mu^2})x^2}=0.
\end{equation}
This yields:
\begin{equation}
\gamma  +i\hbar m \theta(i\mu+\sqrt{\omega^2-\mu^2})=0.
\end{equation}
Let us choose $\gamma$ and $\theta$ such that:
\begin{equation}
 \rm Im \ \theta=0,\quad \rm Re \ \gamma=\theta m \mu, \quad \rm Im \ \gamma=-\theta m \sqrt{\omega^2-\mu^2}.
\end{equation}
It follows using the condition (\ref{norm}) that
\begin{equation}
 \theta=\frac{1}{\sqrt{2m\hbar}(\omega^2-\mu^2)^{\frac{1}{4}}}.
\end{equation}
Consequently, the annihilation and creation operators read:
\begin{eqnarray}
 \hat a&=&\frac{1}{\sqrt{2m\hbar}(\omega^2-\mu^2)^{\frac{1}{4}}}\Bigl[m (i\mu+\sqrt{\omega^2-\mu^2}) \hat x+i\hat p\Bigr],\\
 \hat a^\dag&=&\frac{1}{\sqrt{2m\hbar}(\omega^2-\mu^2)^{\frac{1}{4}}}\Bigl[m (-i\mu+\sqrt{\omega^2-\mu^2}) \hat x-i\hat p\Bigr].
\end{eqnarray}
Equivalently, we can write
\begin{eqnarray}
 \hat p&=&-i\sqrt{\frac{m\hbar}{2\sqrt{\omega^2-\mu^2}}}\Bigl[\sqrt{\omega^2-\mu^2}(\hat a-\hat a^\dag)-i\mu (\hat a+\hat a^\dag)\Bigr] ,\label{pp1}\\
 \hat x&=&\sqrt{\frac{m\hbar}{2\sqrt{\omega^2-\mu^2}}}(\hat a+\hat a^\dag).\label{pp2}
\end{eqnarray}
\section{Variances}
Let us denote by $|n\rangle$ the wave function in the occupation numbers space, such that $\langle x|n\rangle=\psi_n(x)$ and  $\hat a^\dag a|n\rangle=n|n\rangle$. It can be easily verified that $\langle n|\hat p|n\rangle=\langle n|\hat x|n\rangle=0$.

By using equation (\ref{pp1}) it follows that 
\begin{equation}
 \langle n|\hat p^2|n\rangle=\frac{\hbar m \omega^2}{2 \sqrt{\omega^2-\mu^2}}(2n+1).
\end{equation}
Similarly, we obtain from equation (\ref{pp2}):
\begin{equation}
 \langle n|\hat x^2|n\rangle=\frac{\hbar}{2 m \sqrt{\omega^2-\mu^2}}(2n+1).
\end{equation}
Defining the variances of $x$ and $p$ by
\begin{equation}
 \Delta^2_n p=\langle n|\hat p^2|n\rangle-\langle n|\hat p|n\rangle^2, \quad \Delta^2_n x=\langle n|\hat x^2|n\rangle-\langle n|\hat x|n\rangle^2,
\end{equation}
we find the uncertainty product:
\begin{equation}
 \Delta_n x \Delta_n p=\frac{\hbar \omega}{\sqrt{\omega^2-\mu^2}}\bigl(n+\frac{1}{2}\bigr).
\end{equation}
The lower bound of the latter product corresponds to $n=0$, that is to the ground state, and we have in general
\begin{equation}
 \Delta x \Delta p\ge\frac{\hbar \omega}{2\sqrt{\omega^2-\mu^2}}.
\end{equation}
This can be put into the following form
\begin{equation}
 \Delta x \Delta p\ge\Bigl(\frac{\hbar}{2}\Bigr)\frac{1}{\sqrt{1-\frac{\mu^2}{\omega^2}}}.
\end{equation}
The right-hand side grows  as $\mu$ approaches $\omega$. All the values within the interval $(\frac{\hbar}{2}, \hbar/2\sqrt{1-\frac{\mu^2}{\omega^2}})$ are thus quantum mechanically forbidden.

Next we check that the coherent states corresponding to the annihilation and creation operators $\hat a$ and $\hat a^\dag$ are effectively minimum-uncertainty states. A coherent state $|\alpha\rangle$ is defined by
\begin{equation}
 \hat a|\alpha\rangle=\alpha |\alpha\rangle,
\end{equation}
where $\alpha$ is a complex number which can be written as $\alpha=|\alpha| e^{i\phi}$. Hence the expectation value of $\hat p$ with respect to the coherent state $|\alpha\rangle$  is given by
\begin{equation}
 \langle \hat p\rangle_\alpha=2\sqrt{\frac{m\hbar}{2\sqrt{\omega^2-\mu^2}}}|\alpha|\Bigl(\sqrt{\omega^2-\mu^2}\sin\phi-\mu\cos\phi\Bigr).
\end{equation}
Similarly, we find that:
\begin{equation}
 \langle \hat p^2\rangle_\alpha=\frac{m\hbar}{\sqrt{\omega^2-\mu^2}}\Bigl((\omega^2-\mu^2)2|\alpha|^2\sin^2\phi+\frac{1}{2})+\mu^2(2|\alpha|^2\cos^2\phi+\frac{1}{2})-2 |\alpha|^2\mu \sqrt{\omega^2-\mu^2} \sin2\phi\Bigr).
\end{equation}
Consequently the variance reads:
\begin{equation}
 \Delta^2_\alpha p=\langle \alpha|\hat p^2|\alpha\rangle-\langle \alpha|\hat p|\alpha\rangle^2=\frac{\hbar \omega^2 m}{2\sqrt{\omega^2-\mu^2}}.
\end{equation}
Moreover, we have
\begin{equation}
 \langle \hat x\rangle_\alpha=2\sqrt{\frac{\hbar}{2m\sqrt{\omega^2-\mu^2}}}|\alpha|\cos\phi,
\end{equation}
and 
\begin{equation}
 \langle \hat x^2\rangle_\alpha=\frac{\hbar}{2m\sqrt{\omega^2-\mu^2}}(2|\alpha|^2\cos2\phi+2|\alpha|^2+1).
\end{equation}
This implies that
\begin{equation}
 \Delta^2_\alpha x=\langle \alpha|\hat x^2|\alpha\rangle-\langle \alpha|\hat x|\alpha\rangle^2=\frac{\hbar}{2m\sqrt{\omega^2-\mu^2}}.
\end{equation}
Hence:
\begin{equation}
 \Delta_\alpha x \Delta_\alpha p=\Bigl(\frac{\hbar}{2}\Bigr)\frac{1}{\sqrt{1-\frac{\mu^2}{\omega^2}}}.
\end{equation}
The latter result confirms that the lower bound of the uncertainty product is attained for the coherent states.

 The expectation value of the position operator $\hat x$ with respect to time-evolved  coherent state $|\alpha(t)\rangle$ is calculated  according to equation (\ref{pp2}) as:
\begin{eqnarray}
 x(t)&=&2\sqrt{\frac{\hbar}{2m\sqrt{\omega^2-\mu^2}}}|\alpha| \Biggl[\cos\phi\Bigl(\cos(t \sqrt{\omega^2-\mu^2})+\frac{\mu}{\sqrt{\omega^2-\mu^2}} \sin(t \sqrt{\omega^2-\mu^2})\Bigr)\nonumber \\ &+&\frac{\sqrt{\omega^2-\mu^2} \sin\phi-\mu \cos\phi}{\sqrt{\omega^2-\mu^2}}\sin(t \sqrt{\omega^2-\mu^2})\Biggr].
 \end{eqnarray}
After some algebra, we obtain that 
\begin{equation}
 x(t)=2\sqrt{\frac{\hbar}{2m\sqrt{\omega^2-\mu^2}}}|\alpha|\Biggl[\cos(t \sqrt{\omega^2-\mu^2}-\phi)+\frac{\mu}{\sqrt{\omega^2-\mu^2}} \sin(t\sqrt{\omega^2-\mu^2}-\phi)\Biggr].
\end{equation}
In a similar way, we find using equation (\ref{pp1}) that the expectation value of the momentum operator $\hat p$ reads:
\begin{equation}
 p(t)=-2\sqrt{\frac{m\hbar}{2\sqrt{\omega^2-\mu^2}}}|\alpha|\Biggl[\mu \cos(t \sqrt{\omega^2-\mu^2}-\phi)+\sqrt{\omega^2-\mu^2} \sin(t\sqrt{\omega^2-\mu^2}-\phi)\Biggr].
\end{equation}
\section{Concluding remarks}
For the present system, it turns out that it is possible to determine exactly the lower bound of the uncertainty product. The latter increases as the parameter $\mu $ approaches the frequency of the oscillator from bellow. The coherent states associated with the constructed annihilation operator  represent effectively minimum-uncertainty states for the system. The expectation values of the position and momentum operators with respect to the time-dependent coherent states resemble the classical solutions of the equations of motion. The Hamiltonian of the system has been investigated in the context of the quantum nonequilibrium dynamics in order to derive the quantum mechanical diffusion coefficients ~\cite{hamd}. It is thus interesting to explore the implications of the form of the uncertainty product on the obtained results. The question of finding the lower bound for thermal states is reported in~\cite{wang}, which could also be explored in the same direction.

\end{document}